\documentclass[preprint,12pt]{elsarticle}

\usepackage{amssymb}
\usepackage{multicol}
\usepackage{graphicx}
\usepackage{booktabs}
\usepackage{bm,mathrsfs,bbm,amscd}
\usepackage[tbtags]{amsmath}
\usepackage{lastpage}
\usepackage{verbatim}

\journal{Physics Review C}

\begin{document}

\begin{frontmatter}



\title{Improved Calculation of the Energy Release in Neutron Induced Fission\tnoteref{label1}}
 \tnotetext[label1]{Corresponding authors}
\author{X.B.Ma\corref{cor1}\fnref{label2}}
 \ead{maxb917@163.com}
\author[label3]{W.L.Zhong}
\ead{zhongwl@ihep.ac.cn}
\author[label2]{L.Z.Wang}
\author[label2]{Y.X.Chen}
\author[label3]{J.Cao}
\address[label2]{North China Electric Power University, Beijing,102206,China \fnref{label2}}
\address[label3]{Institute of High Energy Physics, Chinese Academy of Sciences, Beijing 100049, China \fnref{label3}}
\begin{abstract}
Fission energy is one of the basic parameters needed in the calculation of antineutrino flux from nuclear reactors. Improving the precision of fission energy calculations is useful for current and future reactor neutrino experiments, which are aimed at more precise determinations of neutrino oscillation parameters. In this article, we give new values for fission energies of some common thermal reactor fuel isotopes, with improvements on three aspects. One is more recent input data acquired from updated nuclear databases. The second improvement is a consideration of the production yields of fission fragments from both thermal and fast incident neutrons for each of the four main fuel isotopes. The third improvement is a more careful calculation of the average energy taken away by antineutrinos in thermal fission involving a comparison of antineutrino spectra from different models. The change in calculated antineutrino flux due to the new values of fission energy is about 0.32$\%$,  and the uncertainties of the new values are about 50$\%$ smaller.
\end{abstract}


\end{frontmatter}

\section{Introduction}
Reactor neutrino experiments have always played a critical role in the history of neutrino physics. For example, the Savannah River Experiment\cite{lab1,lab1-1} by Reines and Cowan in 1956 first detected the neutrino. The KamLAND\cite{lab3} experiment confirmed neutrino oscillation and explained the solar neutrino deficit together with the SNO experiment in the first few years after 2000. Just before that, the CHOOZ \cite{lab2} experiment determined the most stringent upper limit of the last unknown neutrino mixing angle, $sin^{2}2\theta_{13}<0.17$ at a 90$\%$ confidence level. After this, a generation of reactor neutrino experiments made efforts to determine the value of $\theta_{13}$. In March of 2012, the Daya Bay collaboration\cite{lab4} discovered a non-zero value for $sin^{2}2\theta_{13}$ at a 5$\sigma$ confidence level, which has fueled discussions about the direction of neutrino physics in the foreseeable future.

The prediction of antineutrino flux and its uncertainty is an indispensable part of reactor neutrino experiments, especially absolute measurement experiments which detect antineutrinos at only a single location. Usually, the following formula is used to calculate the antineutrino flux from one reactor core:
\begin{equation}\label{eq1}
S(E_{\nu})=\frac{W_{th}}{\sum_{i}(f_{i}/F)E_{i}}\sum\limits_{i}(f_{i}/F)S_{i}(E_{\nu})
\end{equation}
where  $W_{th}$ (MeV/s) is the thermal power of the core, $E_{i}$ (MeV/fission) is the energy released per fission for isotope $i$, $f_{i}$ is the fission rate of isotope $i$, and $F$ is the sum of  $f_{i}$ for all isotopes. Thus, $f_{i}/F$ is the fission fraction of each isotope. $S_{i}(E_{\nu})$ is the antineutrino energy spectrum of isotope $i$, which is normalized to one fission. Normally, $W_{th}$ and $f_{i}/F$ of each isotope are supplied by the nuclear power plants of the reactor neutrino experiments. This leaves $E_{i}$ and $S_{i}(E_{\nu})$ as the two decisive parameters for the accurate calculation of antineutrino flux. In this article, we restrict our discussion to $E_{i}$ only. We will explain how we improve the precision of the calculation of $E_{i}$ on three aspects, and compare the new value and its error with those from predecessors.

\section{ Calculation Method of the Energy Release in Fission $E_{f}$}

The energy release per fission $E_{f}$ can be represented as the sum of the four terms\cite{Kopeilin}.
\begin{equation}\label{Ef}
E_{f}=E_{total}-<E_{\nu}>- \Delta E_{\beta\gamma}+E_{nc}
\end{equation}
where $E_{total}$ is the total energy in fission from the instant at which the neutron that induces the process is absorbed to the completion of the beta decays of the product fragments and their transformation into beta-stable atoms. It includes the total kinetic energy of the fission fragments, total kinetic energy of the emitted prompt and delayed neutrons, and all kinetic energy of the emitted photons, $\beta$ particles, and antineutrinos. $<E_{\nu}>$ is the mean energy carried away by antineutrinos that are produced in the beta decay of fission fragments, $\Delta$$E_{\beta\gamma}$ is the energy of beta electrons and photons from fission fragments that did not decay at a given instant of time. $E_{nc}$ is the energy released in neutron capture (without fission) by various materials of the reactor core.

The energy from the fission process that remains in the reactor core and is transformed into heat can be defined as effective fission energy $E_{eff}$:
\begin{equation}\label{Eeff}
E_{eff}=E_{total}-<E_{\nu}>- \Delta E_{\beta\gamma}
\end{equation}
then, relation(\ref{Ef}) can be recast into the form
\begin{equation}\label{Ef1}
E_{f}=E_{eff}+E_{nc}
\end{equation}
If we calculate $E_{total}$, $<E_{\nu}>$,  $\Delta$$E_{\beta\gamma}$, and $E_{nc}$, then we can obtain a value of the energy release in fission, $E_{f}$.

\section{Calculation Procedures and Results }
\subsection{Total fission energy $E_{total}$}
\begin{figure}
  \centering
  \includegraphics[width=9cm]{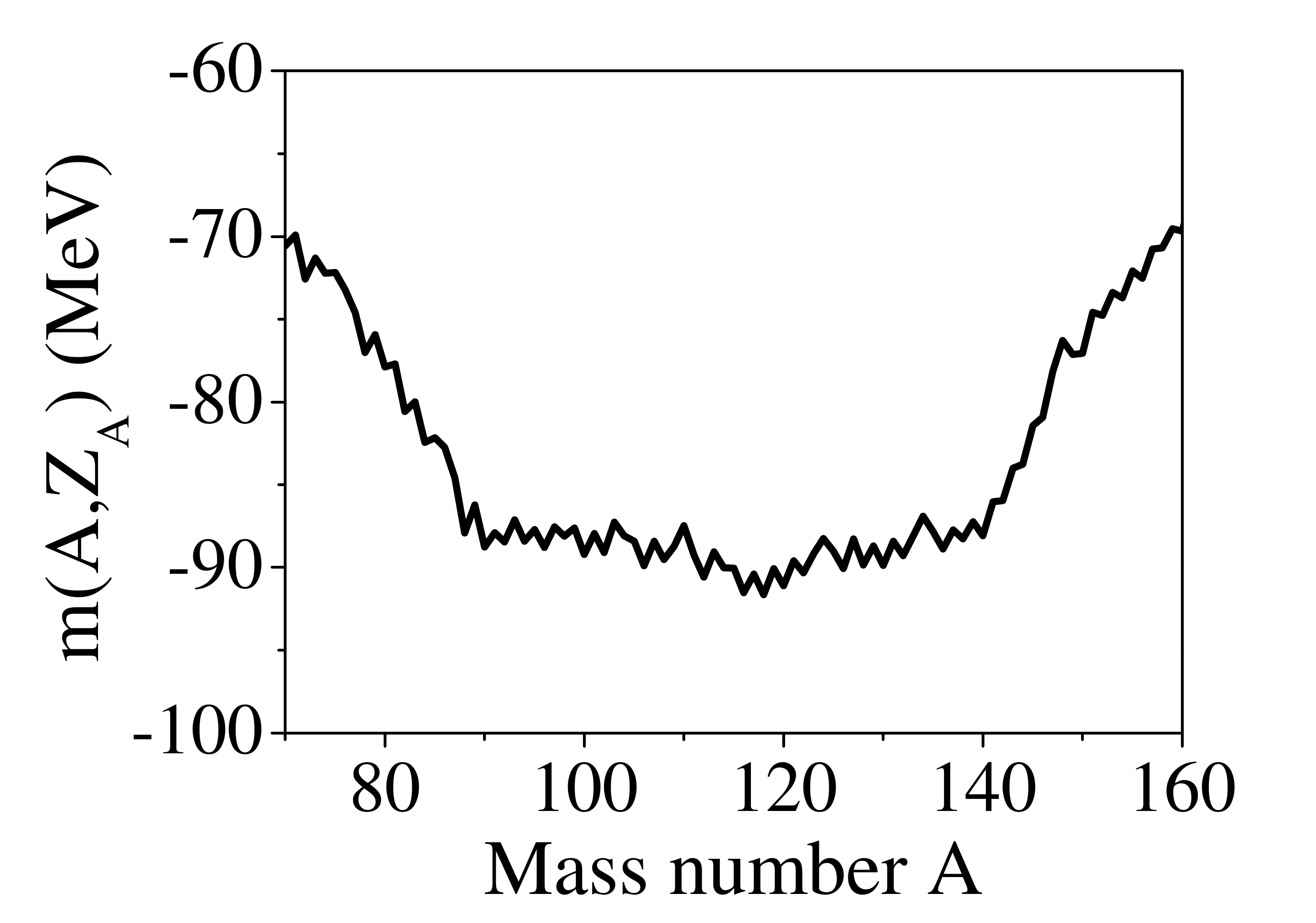}
\caption{ Mass excess $m(A,Z_{A})$ for beta-stable atoms as a function of the mass number A}
\label{massexcess}
\end{figure}

Total fission energy $E_{total}$ can be obtained by directly applying the energy-conservation law. The formula is\cite{Kopeilin}
\begin{equation}\label{eq3}
M(A_{0},Z_{0})+M_{n}=\sum{y_{A}}M(A,Z_{A})+\nu M_{n}+E_{total}
\end{equation}
where $M(A_{0},Z_{0})$ is the atomic mass of the isotope undergoing fission; $A_{0}$ and $Z_{0}$ are its mass and charge numbers respectively; $M_{n}$ is the neutron mass; summation is performed over the mass number A of beta-stable fission products; $M(A,Z_{A})$ is the atomic mass of the product, and $y_{A}$ is its yield, $\Sigma{y_{A}}=2$. The values of A range from 66 through 172. $\nu$ is the mean total number of the prompt and delayed fission neutrons. Using the condition that the number of nucleons is conserved in the fission process and introducing the mass excess for atoms $m(A,Z)$, equation (\ref{eq3}) can be rewritten as:
\begin{equation}\label{eq4}
E_{total}=m(A_{0},Z_{0})-\sum{y_{A}m(A,Z_{A})} - (\nu-1)m_{n}
\end{equation}
where $m(A,Z)=M(A,Z)-Am_{0}$ ($m_{0}$ is one atomic mass unit) and $m_{n}=M_{n}-m_{0}=8.07131710\pm0.00000053$ MeV is the neutron mass excess. Thus, $m(A_{0},Z_{0})$ and $m(A,Z_{A})$ are the mass excess of the isotope undergoing fission and of the fission products, respectively. These values can be obtained from the mass excess evaluation in ATE2003\cite{ATE2003}. Mass excess $m(A,Z_{A})$ for beta-stable atoms is shown in Fig.\ref{massexcess}.

\begin{table}[htbp]
\begin{center}
\caption{Fission ratios of $^{235}$U,$^{238}$U,$^{239}$Pu and $^{241}$Pu induced by thermal and fast neutrons (\%)}
\label{fragment}
\begin{tabular}{cccc}
 \hline
~Fissile isotopes~ &~ ~Thermal neutron~~ &~~ Fast neutron~~&~Error \\
\hline
$^{235}$U  &	76.82 &  23.18 & 0.6 \\
$^{238}$U  &	0.00 &  100.00 & 1.0$\times10^{-7}$ \\
$^{239}$Pu &	90.25 &  9.75 & 0.2 \\
$^{241}$Pu &	83.11 & 16.89 & 0.4 \\
  \hline
\end{tabular}
\end{center}
\end{table}

\begin{figure}
  \centering
  \includegraphics[width=9cm]{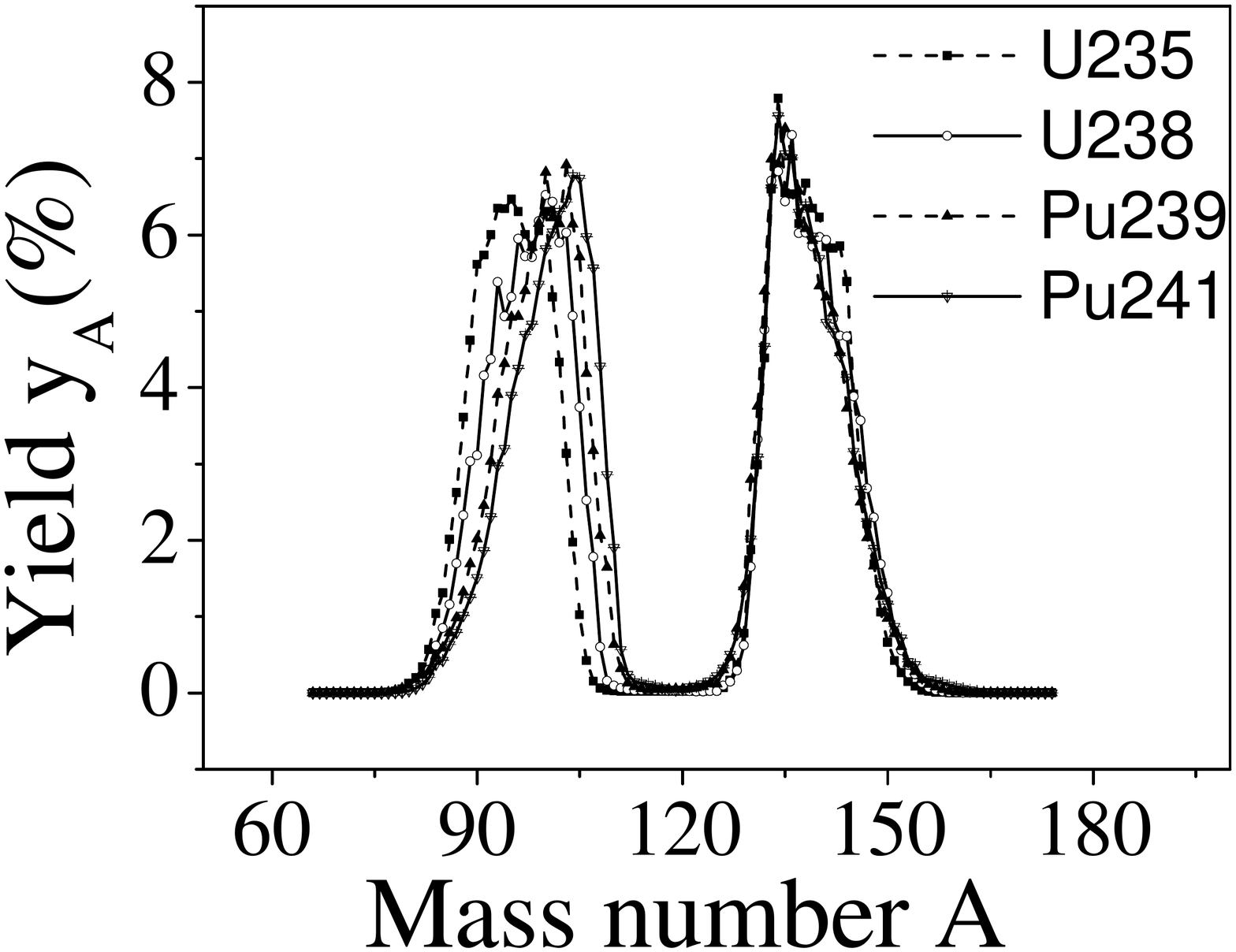}
\caption{Total yield $y_{A}$ of beta-stable fragments from the fission of uranium and plutonium isotopes}
\label{yieldtu}
\end{figure}

According to the INDC\cite{yadis} and other nuclear databases\cite{endf-b7}, for each isotope, the yield $y_{A}$ of each fission fragment in thermal neutron induced fission is different from that of the same fission fragment in fast neutron induced fission. Up to now, calculations have simply treated all fissions of $^{238}$U as being induced by fast neutrons and all fissions of $^{235}$U,$^{239}$Pu, and $^{241}$Pu as being induced by thermal neutrons. However, reactor core simulation data from the Daya Bay Nuclear Power Plant shows that some fissions of $^{235}$U,$^{239}$Pu, and $^{241}$Pu are also induced by fast neutrons. The average fission ratios of the four isotopes from thermal neutrons and fast neutrons during reactor stable running times are shown in table \ref{fragment}. In our calculation, we obtain the thermal fission yield $y_{A_{t}}$ (with error) and fast fission yield $y_{A_{f}}$ (with error) of each fission fragment directly from the INDC database. To include the fission processes from both thermal neutrons and fast neutrons, we use the ratios in table \ref{fragment} to weight $y_{A_{t}}$ and $y_{A_{f}}$ to obtain the average yield $y_{A}$ for each fission fragment of each isotope. The results of $y_{A}$ are shown in Fig.\ref{yieldtu}. The mean total number of the emitted prompt and delayed fission neutrons $\nu$ (with errors) are also obtained directly from the INDC database\cite{yadis}. The precision of $\nu$ from the database is far better than that obtained from a calculation using nucleon number conservation ($A_{0}$ + 1 = $\sum{y_{A}A}$ + $\nu$), which gives a relative error of $\nu$ up to 90$\%$ after error propagation.

With information from the latest nuclear databases, including mass excess, fission yield $y_{A}$, and mean fission neutron number, the total fission energy $E_{total}$ of each isotope is obtained by taking into account both thermal and fast incident neutrons. The parameter values from the latest databases and $E_{total}$ results are in Table \ref{etotal}.\\

\begin{table}[htbp]
\caption{Parameters and $E_{total}$ of $^{235}$U,$^{238}$U,$^{239}$Pu and $^{241}$Pu}
\label{etotal}
\begin{tabular}{cccccc}
\hline
 Fissile  & Mass excess   &                       &Fission &                          &   \\
 isotopes & $m(A_{0},Z_{0})$  & $\sum y_{A}m(A,Z_{A})$ &  neutrons $\nu$ & ($\nu$-1)$m_{n}$  & $E_{total}$  \\
 \hline
$^{235}U$  &40.9205$\pm$0.0018 &-173.859$\pm$0.062  &  2.4355$\pm$0.0023 & 11.586$\pm$0.019 &203.19$\pm$0.06  \\
$^{238}$U  &47.3089$\pm$0.0019 &-173.687$\pm$0.058  &  2.819$\pm$0.020 & 14.682$\pm$0.161 &206.32$\pm$0.17  \\
$^{239}$Pu &48.5899$\pm$0.0018 &-174.196$\pm$0.060  &  2.8836$\pm$0.0047 & 15.203$\pm$0.038 &207.58$\pm$0.07  \\
$^{241}$Pu &52.9568$\pm$0.0018 &-174.100$\pm$0.070  &  2.9479$\pm$0.0055 & 15.722$\pm$0.044 &211.33$\pm$0.08  \\
\hline
\end{tabular}
\end{table}

\subsection{ Average antineutrino energy $<E_{\nu}>$, and $\Delta E_{\beta\gamma}$}
\begin{figure}
  \centering
\includegraphics[width=15cm]{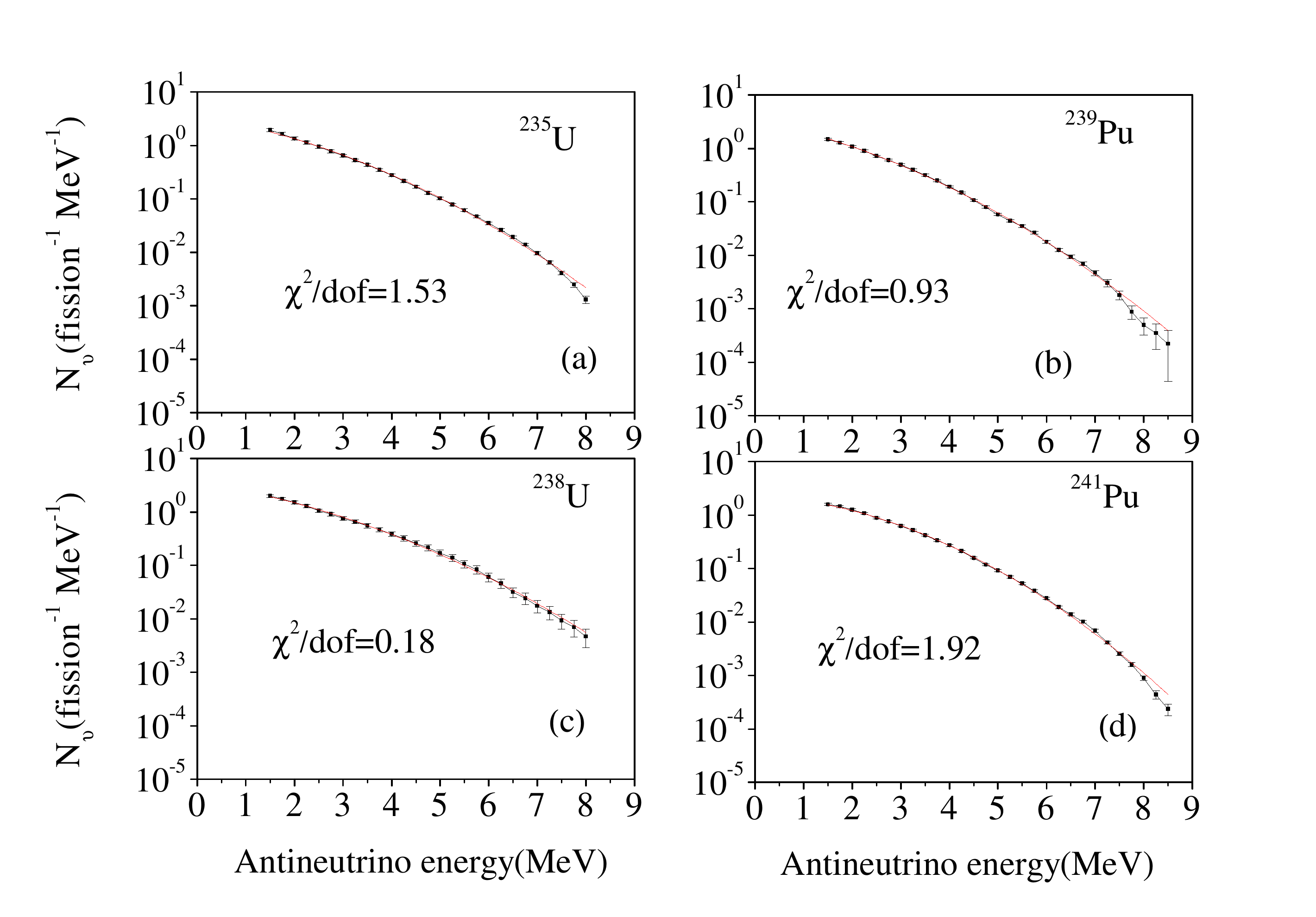}
\caption{Reactor antineutrino spectra of $^{235}$U,$^{238}$U,$^{239}$Pu and $^{241}$Pu}
\label{u5891fit}
\end{figure}

To estimate the average energy of antineutrinos from the fission fragments of $^{235}$U,$^{239}$Pu and $^{241}$Pu, we use the beta-to-antineutrino conversion spectra of the Laue-Langevin Institute (ILL) \cite{ill1,ill2,ill3}. The errors of these spectra are from ILL beta spectra measurements and ILL beta-to-antineutrino conversion method\cite{ill1, ill2}. In the case of $^{238}$U, we use the theoretical $\bar{\nu}_{e}$ spectrum from P. Vogel, et al. \cite{u238spa}, where the errors are theoretically estimated. We calculate non-equillibrium corrections and apply them to the ILL spectra. \\
To determine $<E_{\nu}>$ for each isotope, the function $y=exp(B_{0}+B_{1}x+B_{2}x^{2})$ is used to fit each spectrum, all of which are limited to neutrino energies above 1.5 MeV. In the function, $y$ is the neutrino number per fission per MeV, $x$ is the neutrino energy, and $B_{0}$, $B_{1}$ and $B_{2}$ are fitting parameters. Fitting results are shown in Fig.\ref{u5891fit}.  The $\chi^{2}/dof$ of the fits to the ILL spectra are all close to one, which shows that the function can describe the antineutrino spectra very well.
Fitting parameters are summarized in Table \ref{fittingpar}. The fits are used to smoothly extrapolate to energies below 1.5 MeV.

\begin{table}[htbp]
\begin{center}
\caption{Values of the fitting parameters}
\label{fittingpar}
\begin{tabular}{ccccc}
 \hline
 Parameter& $^{235}$U & $^{238}$U&  $^{239}$Pu & $^{241}$Pu \\
\hline
  $B_{0}$      &1.25636	& 1.26119  &1.20114 & 0.87170  \\
 $B_{1}$   &-0.33897 &  -0.30588& -0.40981& -0.13055  \\
 $B_{2}$   &-0.007309& -0.06253&-0.07690 & -0.10355 \\
  \hline
\end{tabular}
\end{center}
\end{table}

\begin{figure}
 \centering
\includegraphics[width=6.7cm]{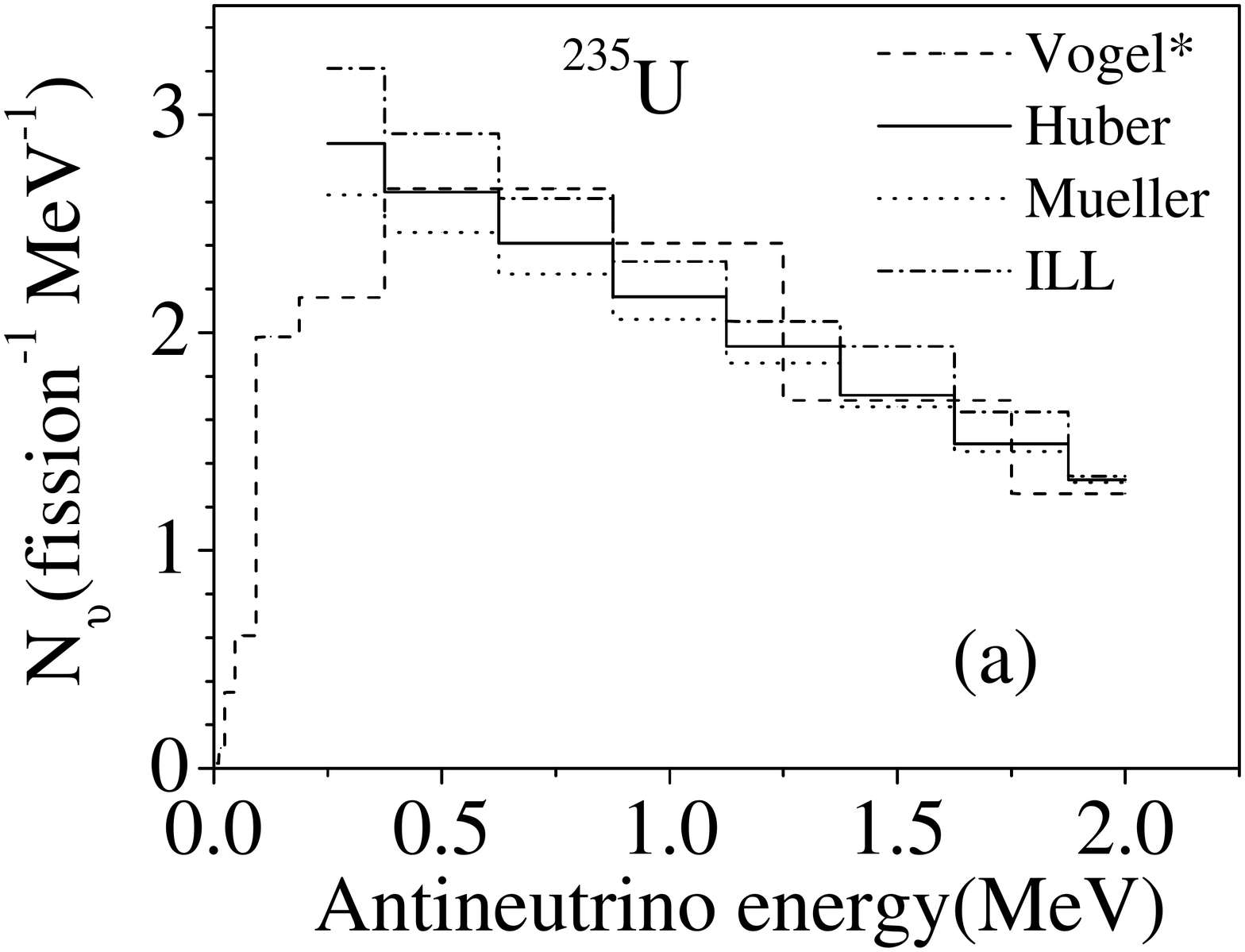}
\includegraphics[width=6.7cm]{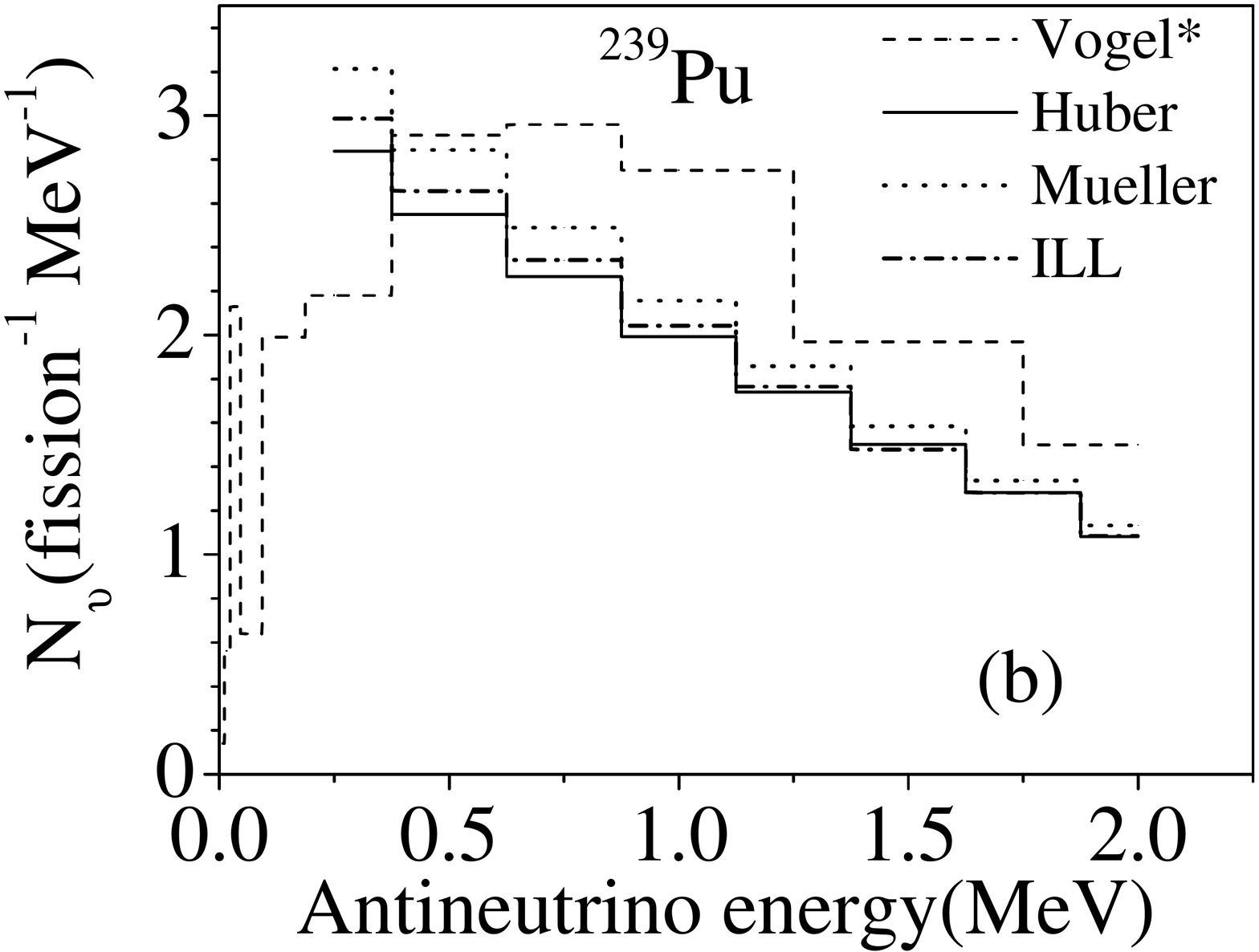}
\includegraphics[width=6.5cm]{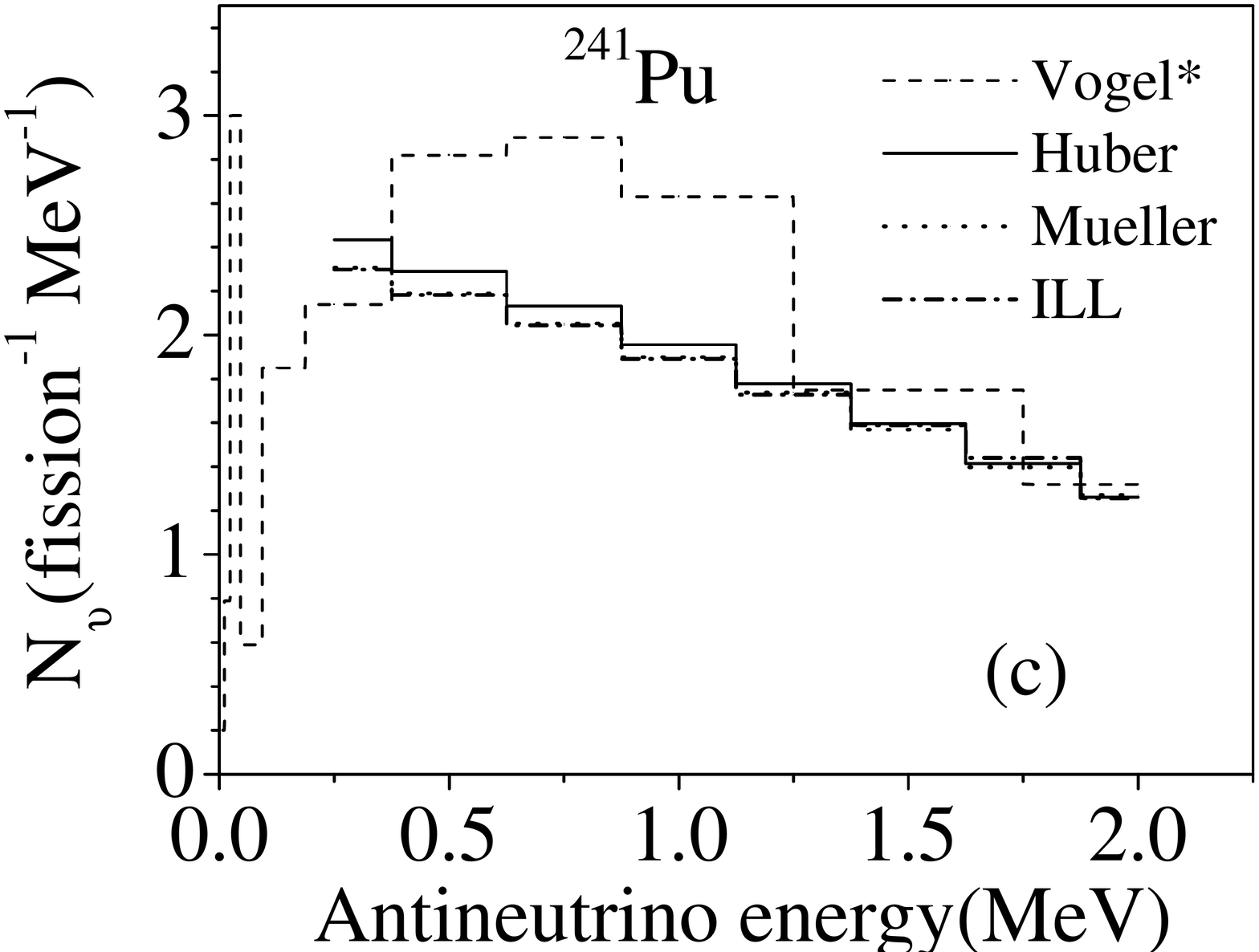}
\includegraphics[width=6.5cm]{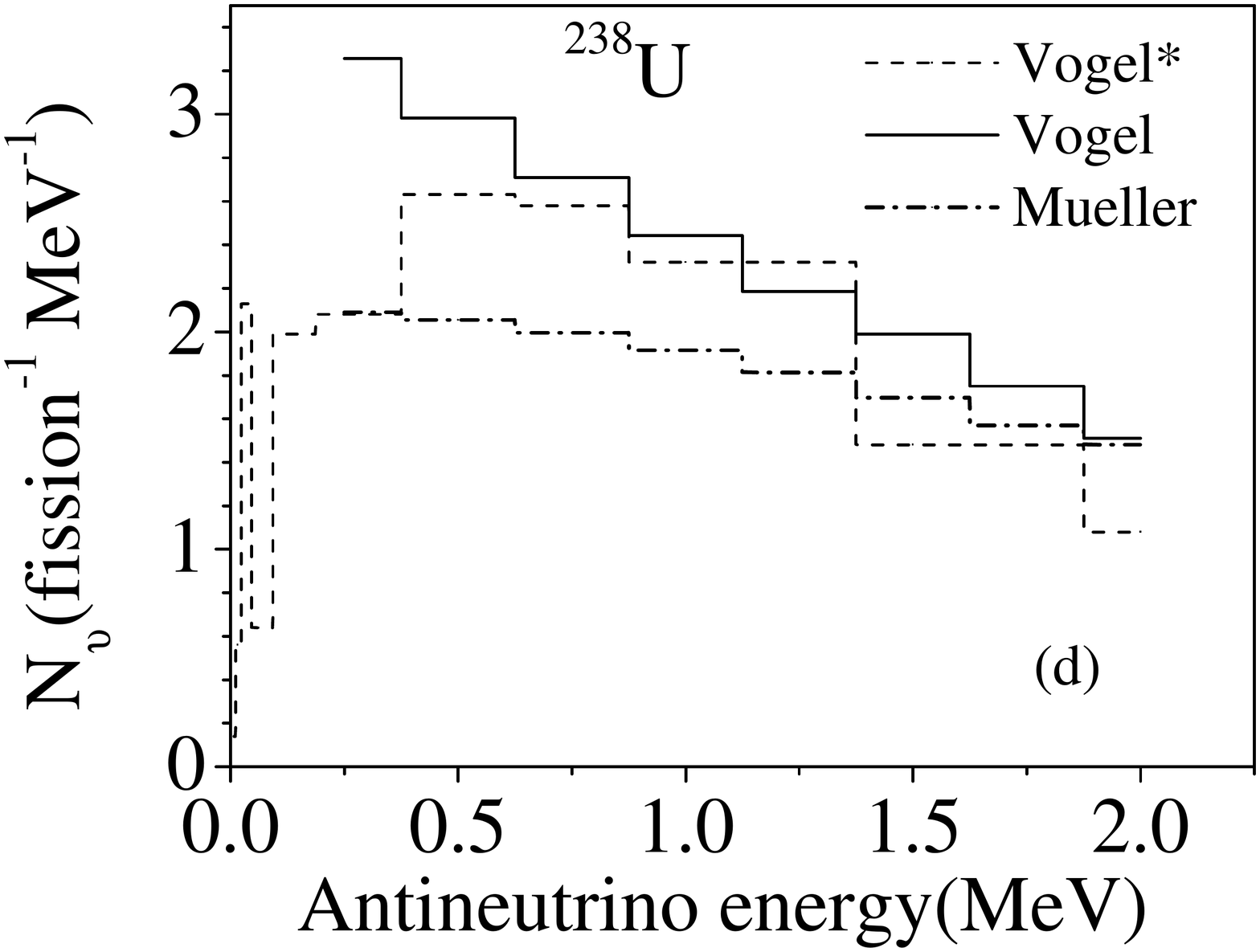}
\caption{Comparison of antineutrino spectra given by different models, ILL, Huber and Mueller et. al. , and theoretical calculation. (a) is spectra of $^{235}$U, (b) is spectra of$^{239}$Pu, (c) is spectra of $^{241}$Pu and, (d) is spectra of $^{238}$U.}
\label{spectraCom}
\end{figure}

Besides the antineutrino spectra from ILL, we also use the beta-to-antineutrino conversion spectra of P. Huber\cite{huber} and Mueller et. al.\cite{mueller}.  We use the same exponential function to fit their isotope spectra and extrapolate to below 2.0 MeV, which is the lower limit of the data. To examine the extrapolation quality, we compare the spectra below 2.0 MeV to the theoretical spectra calculated by P. Vogel\cite{vogel2}.  Fig.\ref{spectraCom}  shows the theoretical spectra from Vogel and the fits of the ILL, Huber and Mueller models, and Vogel's spectra. For antineutrinos above 2.0 MeV, Table \ref{antiE_gt2} summarizes the results from different models ($^{235}U$, $^{239}Pu$ and $^{241}Pu$) or theoretical calculations ($^{238}U$) of antineutrino spectra for each isotope, and the average energies and errors. For antineutrinos below 2.0 MeV, Table \ref{antiE_lt2} summarizes the results from different models and theoretical calculations of antineutrino spectra for each isotope, and the average energies and errors. As one can see from table \ref{antiE_gt2} and table \ref{antiE_lt2}, the main source of the errors of the average energies of antineutrinos is from low energies, below 2.0 MeV.

\begin{table}[htbp]
\begin{center}
\caption{Average energy taken away by antineutrinos above 2.0 MeV.}
\label{antiE_gt2}
\begin{tabular}{cccccc}
\hline
 Fissile   & $<E_{\nu}>$   &$<E_{\nu}>$ & $<E_{\nu}>$ & $<E_{\nu}>$ &  Average  \\
 isotopes   & Huber & Mueller & ILL & Vogel & (MeV) \\
 \hline
$^{235}$U  & 5.292    &   5.259   &  5.126   & - &  5.226$\pm$0.051  \\
$^{238}$U  &  -      &    7.366   &  -   & 6.714 &  7.040$\pm$0.326 \\
$^{239}$Pu & 3.840   &   3.824    &   3.733  & - & 3.799$\pm$0.033  \\
$^{241}$Pu & 5.019   &    4.990   &   4.859  & - & 4.956$\pm$0.049  \\
\hline
\end{tabular}
\end{center}
\end{table}

\begin{table}[htbp]
\begin{center}
\caption{Average energy taken away by antineutrinos below 2.0 MeV.}
\label{antiE_lt2}
\begin{tabular}{ccccccc}
\hline
 Fissile  &$<E_{\nu}>$ & $<E_{\nu}>$ & $<E_{\nu}>$  & $<E_{\nu}>$  & $<E_{\nu}>$ &   Average   \\
 isotope & Vogel$^{\ast}$ & Huber & Mueller & ILL & Vogel$^{\dagger}$ & (MeV) \\
 \hline
$^{235}$U  & 4.013   &   3.713   &  3.561    & 4.034  & - & 3.830$\pm$0.100  \\
$^{238}$U  &  3.730   &   -   & 3.477  & - & 4.225 &  3.810$\pm$0.220 \\
$^{239}$Pu & 4.130    &   3.341   &   3.588   & 3.390 & - & 3.612$\pm$0.181  \\
$^{241}$Pu & 3.801   &    3.396   &   3.319   &  3.334 & - & 3.462$\pm$0.114  \\
\hline
\end{tabular}
\end{center}
${^{\ast}~ results~ of~ the~ theoretical~ spectra~ of~ the~four~isotopes ~ below~ 2.0~MeV} $\\
${^{\dagger}~fit~ of~ ^{238}U~theoretical~spectrum~below~2.0~MeV}$
\end{table}

The total average energy carried by antineutrinos $<E_{\nu}>$ after summing the portions above and below 2.0 MeV is shown in Table \ref{eneutrino}. The fit result for the theoretical spectrum of $^{238}$U is also included. The results are consistent with those in reference \cite{Kopeilin}.

\begin{table}[htbp]
\begin{center}
\caption{Average energy carried away by antineutrinos}
\label{eneutrino}
\begin{tabular}{ccc}
 \hline
Fissile  & $<E_{\nu}>$   & \cite{Kopeilin}   \\
 isotopes   & (MeV) & (MeV) \\
 \hline
$^{235}$U  &	 9.06$\pm$ 0.13 &  9.07$\pm$0.32  \\
$^{238}$U  &	 10.85$\pm$ 0.39 &  11.00$\pm$0.80   \\
$^{239}$Pu &	 7.41$\pm$ 0.18  &  7.22$\pm$0.27   \\
$^{241}$Pu &	 8.42$\pm$ 0.12  &  8.71$\pm$0.30  \\
 \hline
\end{tabular}
\end{center}
\end{table}

The kinetic energies of $\beta$ particles and photons from the complete beta decay of  fission fragments are part of the total fission energy, $E_{total}$. However, at the instant of observation, the decay processes of some long-life isotopes have not yet completed. The correction of $\Delta E_{\beta\gamma}$ is used to subtract the kinetic energies of $\beta$ particles and gammas which have not been emitted. Values for $\Delta E_{\beta\gamma}$ are taken from \cite{Kopeilin} as shown in Table \ref{ebatagama}. These values correspond to a fuel irradiation time in the middle of the standard operating period of a pressurized water reactor.

\begin{table}[htbp]
\begin{center}
\caption{ Energy $\Delta E_{\beta\gamma}$}
\label{ebatagama}
\begin{tabular}{cc}
 \hline
Fissile isotopes &  $\Delta E_{\beta\gamma}$ (MeV) \\
\hline
$^{235}$U  &	0.35$\pm$ 0.02  \\
$^{238}$U  &	0.33$\pm$ 0.03  \\
$^{239}$Pu &	0.30$\pm$ 0.02   \\
$^{241}$Pu &	0.29$\pm$ 0.03   \\
 \hline
\end{tabular}
\end{center}
\end{table}

\subsection{Energy released in neutron capture $E_{nc}$}
In addition to the energy released directly in the fission process, some energy is released in neutron capture upon reactor materials. The total amount of this energy depends on the composition of the reactor materials and the probability of neutron absorption on these materials, therefore it also varies with fuel burning time. In the middle of the reactor operating period, the energy from neutron capture processes $E_{nc}$ which converts into thermal energy in fuel isotope $i$ can be described:
\begin{equation}\label{eqEnc}
E_{nc}=(\bar{\nu_{i}}-1)\bar{Q}
\end{equation}
where $\bar{Q}$ is the mean energy released per capture and $\bar{\nu_{i}}$ is the average number of emitted neutrons per fission. In reference\cite{james1},  a $\bar{Q}$ of $6.1 \pm 0.3$ MeV is given by simply considering a wide range of reactor material compositions. In reference \cite{Kopeilin}, the probabilities of the absorption of neutrons by various materials, and the time evaluations of fuels during burn-up periods are both considered. The average value of $5.97 \pm 0.15$ MeV per neutron capture is given for the middle of the reactor operation period. Its variation within the time interval from one day to the end of the operating period is about 0.55 MeV. In this paper, we use the $\bar{Q}$ value from reference \cite{Kopeilin} and $\bar{\nu_{i}}$ from INDC, which was mentioned earlier for the calculation of $E_{total}$. The results of $E_{nc}$ are shown in Table \ref{Enc}.

\begin{table}[htbp]
\begin{center}
\caption{ Neutron capture released energy $E_{nc}$}
\label{Enc}
\begin{tabular}{cc}
 \hline
Fissile isotopes &  $E_{nc}$ (MeV) \\
\hline
$^{235}$U  &	8.57 $\pm$ 0.22 \\
$^{238}$U  &	10.86 $\pm$ 0.30 \\
$^{239}$Pu &	11.25 $\pm$ 0.28 \\
$^{241}$Pu &	11.63 $\pm$ 0.29 \\
 \hline
\end{tabular}
\end{center}
\end{table}

\subsection{Energy release per fission}
The energy release per fission is required for reactor antineutrino flux calculations, and is usually defined without the kinetic energy of the incident and emitted neutrons\cite{Kopeilin,james1}. However, in the WIMS-D formatted libraries\cite{wimsd} and reference \cite{mufnumber}, the energy release per fission includes the contributions from the kinetic energy of incident neutrons and from the decay of the capture products:
\begin{equation}\label{Ef2}
E_{f}=E_{eff}+E_{nc}+E_{in}
\end{equation}
where $E_{in}$ is the kinetic energy of incident neutrons.  For one isotope, at each step of its fission chain, an amount of energy from the emitted fission neutrons has to be used as the incident neutron energy for the next step in the fission chain. Therefore, the amount energy of $E_{in}$ can not transform into heat until the end of the fission chain.  From the view of energy conservation, at the end of the fission chain, $E_{f}=E_{eff}+E_{nc}+E_{in}$. However, as long as the fission chain has not reached its end, $E_{in}$ has not converted into heat, and therefore has not contributed to the reactor thermal power. Thus, $E_{f}$ should be equal to $(E_{eff}+E_{nc})$. Our calculations of $E_{f}$ without $E_{in}$, and of $E_{eff}$ are shown in Table \ref{efission}, along with $E_{f}$ from reference\cite{Kopeilin}. For $^{239}Pu$, the $E_{f}$ directly stated in reference \cite{Kopeilin} is 209.99 MeV, which should be the sum of  $E_{eff}$ and $E_{nc}$. According to the values of $E_{eff}$ and $E_{nc}$ in the same reference, $E_{f}$ of $^{239}Pu$ should be 210.99 MeV. Thus, we list 210.99 MeV in table \ref{efission}. As one can see in the table, the new $E_{f}$ values are systematically a little larger than those in reference \cite{Kopeilin} and the new errors are about $50\%$ smaller. The contributions to the improved errors of $E_{f}$ are from the calculations of $E_{total}$ and $<E_{\nu}>$.

\begin{table}[htbp]
\begin{center}
\caption{ Energy release per fission}
\label{efission}
\begin{tabular}{cccccc}
 \hline
Fissile      &  $E_{eff}$ & $E_{f}$\cite{Kopeilin}&  $E_{f}$ \\
  isotopes   &      (MeV)    &     (MeV)                &   (MeV)   \\
\hline
$^{235}$U   &193.79$\pm$0.14 & 201.92$\pm$0.46 & 202.36$\pm$0.26  \\
$^{238}$U   &195.13$\pm$0.43 & 205.52$\pm$0.96 & 205.99$\pm$0.52  \\
$^{239}$Pu  &199.87$\pm$0.20 & 210.99$\pm$0.60 & 211.12$\pm$0.34   \\
$^{241}$Pu  &202.63$\pm$0.15 & 213.60$\pm$0.65 & 214.26$\pm$0.33 \\
 \hline
\end{tabular}
\end{center}
\end{table}

\section{Impact on the Antineutrino Flux}
\begin{figure}
\begin{center}
\includegraphics[width=9cm]{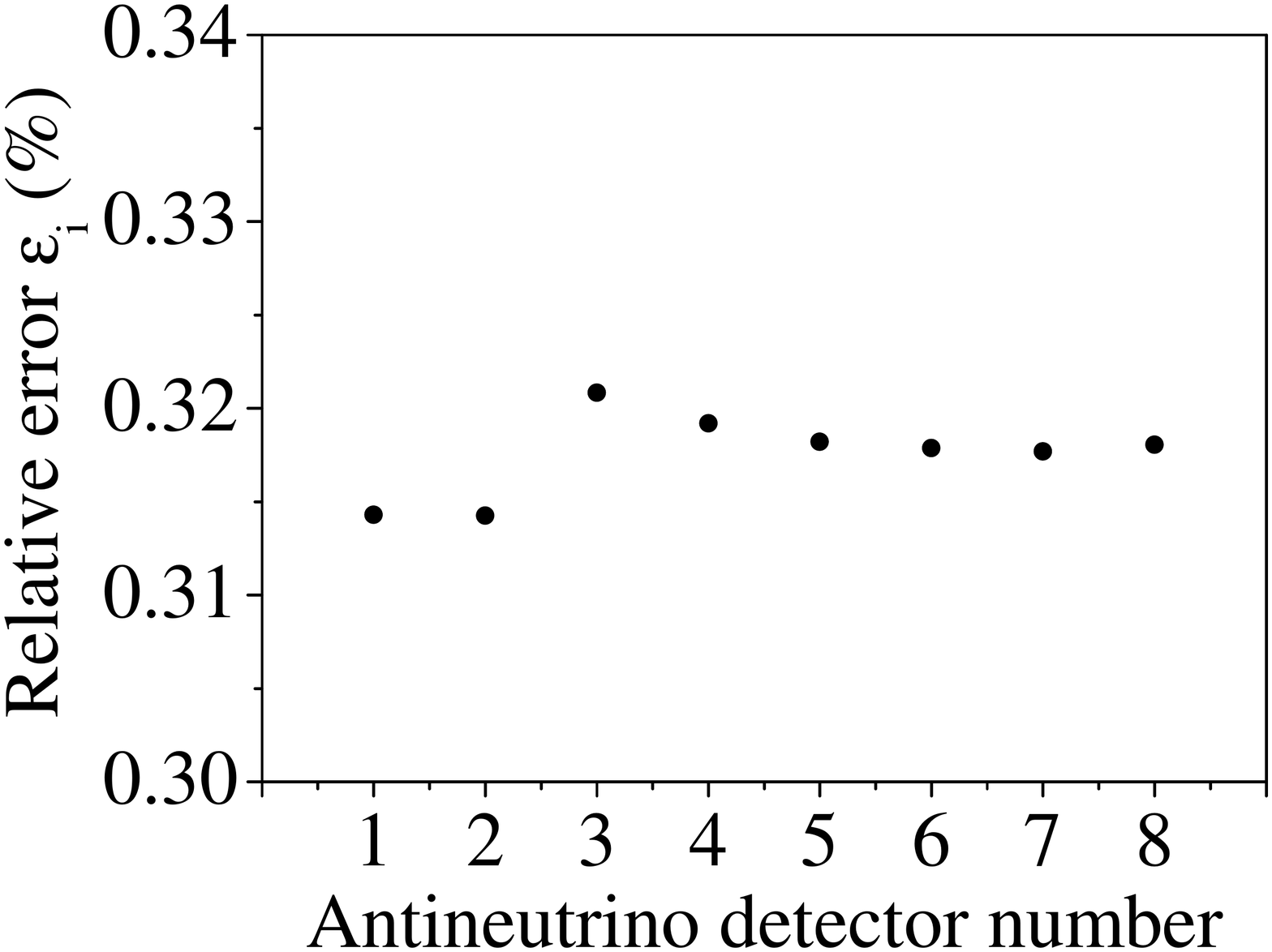}
\caption{Change of average weekly antineutrino flux with the input of old and new values of fission energy}
\label{relative}
\end{center}
\end{figure}

To quantify the effect of the new values for energy per fission on antineutrino flux expectation in a reactor neutrino experiment, we use reactor data from the Daya Bay experiment to calculate the expected average weekly antineutrino flux at the eight antineutrino detectors. The flux obtained with the input of our new fission energy values is denoted as $\phi_{i}$, and that obtained with the values in reference \cite{Kopeilin} is denoted as  $\phi^{'}_{i}$. We define the relative error $\varepsilon_{i}$ as
 \begin{equation}\label{relativeerr1}
\varepsilon_{i}=|\phi^{'}_{i}-\phi_{i}|/\phi_{i}
\end{equation}
where $i$ is the antineutrino detector number. The relative error of the weekly average antineutrino flux detected at each detector $\varepsilon_{i}$ is shown in Fig. \ref{relative}. The first two antineutrino detectors are at one near experimental site, called the Daya Bay site. The third and fourth detectors are located at the other near site, called the Ling Ao site. The remaining four detectors are at the far site. Each detector receives antineutrinos from three reactor pairs. Due to the differences in fission fractions of isotopes between different reactor cores and the differences in baselines between detectors and reactors, the relative error varies among detectors, but they are all around $0.32\%$. The neutrino flux calculated with the new values is a little smaller because of a larger average energy release per fission.

\section{Conclusion}
To improve the precision of the calculation of the energy release in fission, we have employed the most recent data from nuclear databases, such as mass excess, yield of fission fragments, and average fission neutron yields. When we apply the yield values to fission fragments, we consider the thermal and fast neutron induced fissions separately, weight them and then sum them. These two considerations help to reduce the uncertainties of $E_{total}$ by about $50\%$ compared to those given by Kopeikin\cite{Kopeilin}. In the calculation of $<E_{\nu}>$, we compare the antineutrino spectra of different models and use the average of different models as the final average energy carried by antineutrinos. For the other two components, $\Delta E_{\beta\gamma}$, and $E_{nc}$, $\Delta E_{\beta\gamma}$ is imported from reference\cite{Kopeilin}, the calculation of $E_{nc}$ uses data of average fission neutron yield from the INDC database\cite{yadis}, and the the estimate of $<E_{\nu}>$ is from our own fitting, which has similar but smaller uncertainties. Adding the four components together, we obtain the final fission energies for $^{235}$U,$^{238}$U,$^{239}$Pu and $^{241}$Pu. They are systematically a little larger than Kopeikin's results\cite{Kopeilin}, with an improvement in uncertainty of about $50\%$. The impact of the new values to the expected antineutrino flux is at the level of $0.3\%$. We also noticed that the differences in fission energy values between reference\cite{Kopeilin,james1} and WIMS-D formatted libraries\cite{wimsd} and reference \cite{mufnumber} are from different treatments of incident neutron kinetic energy $E_{in}$. Considering that the incident neutron kinetic energy is used to propagate the fission chain and will not convert into reactor heat until the end of the fission chain, we do not include the incident neutron kinetic energy into the fission energy when calculating antineutrino flux.

\section*{Acknowledgements}
We appreciate the support from members of the Daya Bay collaboration, particularly the enlightening discussions with P. Huber and contributions from Logan Lebanowski. This work was supported by National Natural Science Foundation of China (No. 11175201), the Fundamental Research Funds for the Central Universities (12MS63).





\end{document}